\documentclass[prl,twocolumn,superscriptaddress,showpacs,nofootinbib,longbibliography]{revtex4-2}

\usepackage[dvips]{graphicx}
\usepackage{amssymb}
\usepackage{amsmath}
\usepackage{natbib}
\usepackage{xcolor}
\usepackage{chngcntr}
\usepackage{enumitem}
\usepackage{hyperref}

\hypersetup{
colorlinks=false,
linktocpage=true,
citebordercolor=[rgb]{0.5,0.5,1.0},
linkbordercolor=[rgb]{1.0,0.5,0.5},
urlbordercolor=[rgb]{0.5,0.5,1.0}
}

\begin{document}

\title{Exact Solution to the Quantum and Classical Dimer Models on the Spectre Aperiodic Monotiling}

\author{Shobhna Singh}
  \affiliation{School of Physics and Astronomy, Cardiff University, The Parade, Cardiff CF24 3AA, United Kingdom}
  \affiliation{H.~H.~Wills Physics Laboratory, Tyndall Avenue, Bristol, BS8 1TL, United Kingdom}
\author{Felix Flicker}
  \affiliation{H.~H.~Wills Physics Laboratory, Tyndall Avenue, Bristol, BS8 1TL, United Kingdom}

\begin{abstract}
The decades-long search for a shape that tiles the plane only aperiodically under translations and rotations recently ended with the discovery of the `spectre' aperiodic monotile. In this setting we study the dimer model, in which dimers are placed along tile edges such that each vertex meets precisely one dimer. The complexity of the tiling combines with the dimer constraint to allow an exact solution to the model. The partition function is $\mathcal{Z}=2^{N_{\textrm{Mystic}}+1}$ where $N_{\textrm{Mystic}}$ is the number of `Mystic' tiles. We exactly solve the quantum dimer (Rokhsar Kivelson) model in the same setting by identifying an eigenbasis at all interaction strengths $V/t$. We find that test monomers, once created, can be infinitely separated at zero energy cost for all $V/t$, constituting a deconfined phase in a 2+1D bipartite quantum dimer model. 
\end{abstract}

\maketitle 

%
%

The dimer model is one of the oldest models in statistical physics. Given a graph (vertices connected by edges), a `perfect dimer matching' is a set of edges (dimers) such that each vertex connects to precisely one member of the set. The dimer model then considers the set of all perfect matchings. The model characterises a wide range of physical processes including adsorption~\cite{Kasteleyn61,Fisher61,FisherTemperley61,Kasteleyn63,Kasteleyn67}, zero modes in electronic tight binding models~\cite{LH50,Lieb89,SchirmannEA23}, and magnetism, where dimers are used for example in analytic approaches to the Ising model~\cite{Baxter}. The \emph{quantum} dimer model (QDM), also called the Rokhsar Kivelson model, introduces quantum superpositions of dimer placements~\cite{RokhsarKivelson88,MoessnerRaman07}. Originally introduced to capture the physics of resonating valence bond states~\cite{Anderson87} in theories of high-temperature superconductivity~\cite{RokhsarKivelson88}, QDMs are now understood to host a range of exotic phenomena such as quantum spin liquids, topological order, and fractionalisation~\cite{MoessnerRaman07,SavaryBalents17}. They have recently been realised experimentally in programmable quantum simulators~\cite{SatzingerEA21,SemeghiniEA21}. 

The utility of classical dimer models derives in part from an efficient method (the `FKT algorithm') for enumerating perfect matchings developed by Fisher, Kasteleyn, and Temperley~\cite{Kasteleyn61,Fisher61,FisherTemperley61,Kasteleyn63,Kasteleyn67}. The result permits an exact solution to any $N$-vertex planar dimer model in the form of the partition function:
\begin{align}
\label{eq:Z}
\mathcal{Z}_N\left[w\right]=\sum_{\mathcal{M}_i\in\mathcal{M}}\,\,\prod_{e\in \mathcal{M}_i} w\left(e\right).
\end{align}
Here, $\mathcal{M}_i$ is a perfect matching in the set of all perfect matchings $\mathcal{M}$, and $e$ are the edges, of the graph. Setting weight $w=1$ on all edges, $\mathcal{Z}\left[1\right]$ counts the number of perfect matchings. From $\mathcal{Z}$ all thermodynamic functions of state immediately follow. Of particular interest is the free energy per dimer~\cite{Wu06} in an \mbox{$N$-vertex} graph
\begin{align}
\label{eq:f}
f_N\left[w\right]=\frac{1}{N/2}\ln\left(\mathcal{Z}_N\left[w\right]\right).
\end{align}

For certain regular graphs admitting periodic embeddings, $\mathcal{Z}$ has been evaluated analytically~\cite{Kasteleyn63,Kasteleyn67,Wu06}. Owing to the importance of graph connectivity to the behaviour of dimer models, they have recently begun to be studied on infinite graphs with aperiodically ordered planar embeddings~\cite{FlickerEA20,LloydEA21,SinghEA23,SchirmannEA23}. Such graphs, which capture the symmetries of physical quasicrystals~\cite{BaakeGrimm,Senechal}, are irregular, meaning their vertices meet different numbers of edges, typically leading to a large degree of frustration in dimer arrangements. They admit planar embeddings which are long-range ordered, meaning their diffraction patterns feature sharp Bragg peaks~\cite{BaakeGrimm,Senechal}, despite lacking a discrete translational symmetry. Examples include the graph version of the Penrose tiling~\cite{Penrose74,FlickerEA20} and the Ammann-Beenker tiling~\cite{GrunbaumShephard,LloydEA21,SinghEA23}. The long range order often permits analytical results; for example, in a modification of the Ammann Beenker tiling an exact solution to the dimer model can be approximated to arbitrary accuracy using transfer matrices~\cite{LloydEA21}. 

\begin{figure}[h]
\includegraphics[width=\columnwidth]{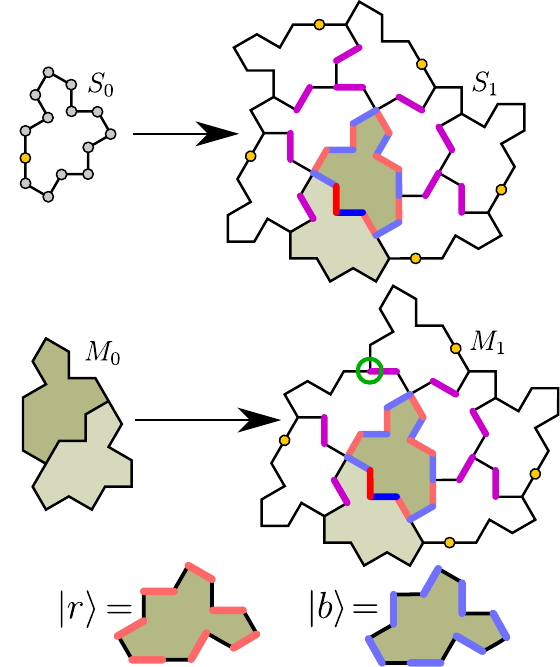}
\caption{The composition rules for the spectre tiling (after Ref.~\onlinecite{spectres}). The vertices of the spectre $S_0$ are indicated; the gold vertex is added whenever it is not implied by the meeting of tiles. The two once-composed tiles $S_1$ and $M_1$ can be pieced together without overlaps to construct the infinite aperiodic tiling. Note that composition mirrors $S_0$ tiles in such a way that only one chirality appears at any level of composition. The Mystic $M_0$ is the two green tiles (the darker tile being the Upper Mystic $M_0^+$). Of the four internal edges of $M_0$ either the red or dark blue dimer must appear in any perfect matching. Choosing red, all pink and purple dimers are forced. Choosing blue, all light blue and purple dimers are forced. The only freedom on internal $S_1$ and $M_1$ edges is therefore two dimer matchings per Mystic (these cases are shown at the bottom, labelled $|r\rangle$ (red) and $|b\rangle$ (blue) for convenience in the quantum model). The gold vertex appears only on the boundaries of $S_1$ and $M_1$, so does not affect this argument.}
\label{fig:implied}
\end{figure}

This year saw a major advance in the study of aperiodic tilings with the discovery of the `spectre' aperiodic monotile~\cite{spectres}. The spectre positively answered the decades-old question of whether there exists a shape that tiles the plane only aperiodically under translations and rotations~\cite{spectres}. Spectre tilings, either finite or infinite, can be created by the `composition rules' in Fig.~\ref{fig:implied}, reproduced from Ref.~\onlinecite{spectres}, in which each tile is replaced with copies of itself so as to build a larger tiling. Each tile becomes its mirror image under composition, meaning that all tiles have the same chirality after each composition. 

Here we provide an exact analytical solution to both the classical and quantum dimer models on spectre tilings. We treat the vertices and edges of the tiles as those of a graph. Since we are only concerned with graph connectivity, we straighten the curved edges of the spectre tiles (resulting in what is termed `Tile(1,1)' in Refs.~\cite{hats,spectres}). Each spectre tile, labelled $S_0$, can have either 13 or 14 edges depending on its environment~\cite{spectres}. To ensure that all tiles are identical at the level of graph connectivity we add a vertex (gold in Fig.~\ref{fig:implied}) to any 13-edge tiles. This makes the graphs bipartite, meaning vertices divide into two sets such that edges only connect vertices in different sets. We discuss the non-bipartite case briefly at the end.

%
%

\emph{Results (classical)} --- Starting from a single spectre $S_0$, a finite number of compositions $\mathcal{N}$ generates a finite connected tile set $S_{\mathcal{N}}$, while an infinite number of compositions results in a tiling of the Euclidean plane~\cite{spectres}. Even though each tile has 14 vertices, the total number of vertices can still be odd; in such cases the number of perfect matchings is zero, since a dimer must connect a pair of vertices. By construction, any tiling built by composition can also be seen as a concatenation of the once-composed tiles $S_1$ and $M_1$ (Fig.~\ref{fig:implied}). The two green spectres in Fig.~\ref{fig:implied} together make up a tile called the `Mystic', which we denote $M_0$. We term the dark green tile the `Upper Mystic' $M_0^+$. It plays a special role as the only entirely internal tile in $S_1$ and $M_1$. It is also marked out as special by appearing $\pi/6$ rotated from the other tiles, which appear only $\pi/3$ rotated amongst themselves (Fig.~\ref{fig:implied}). The Mystic $M_0$ contains four internal edges. Of these four, exactly one of the two central edges must be covered by a dimer in any perfect matching. Either choice forces a range of other dimers. Choosing the red dimer in Fig.~\ref{fig:implied} forces all the pink and purple dimers; instead choosing the blue dimer forces all the light blue and purple dimers. The purple dimers are the same in both cases, and so these edges are always covered in any perfect matching. In fact the figure demonstrates that \textit{every} non-boundary edge of $S_1$ and $M_1$ either lies on $M_0^+$, or has a fixed dimer occupation as shown. The only degrees of freedom are the two dimer matchings per $M_0^+$ (and therefore per $M_0$), or possibly the boundaries of $S_1$ or $M_1$ within larger regions. From now on we focus on $S_\mathcal{N}$ regions unless otherwise stated. 

\begin{figure}[h]
\includegraphics[width=\columnwidth]{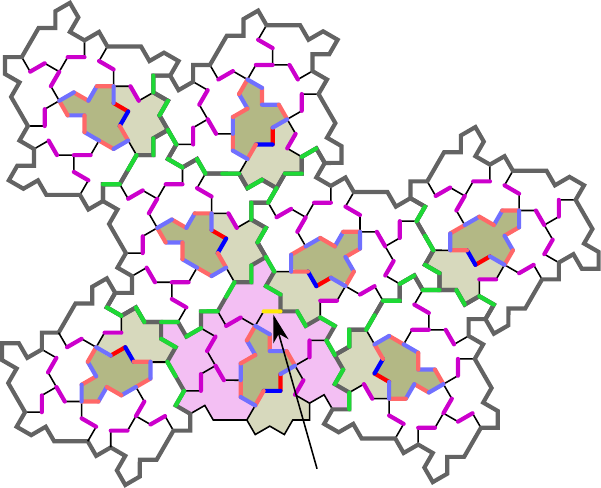}
\caption{Twice-composition of the spectre, $S_2$. The mirror of $M_1$ (Fig.~\ref{fig:implied}) is highlighted in pink. The gold dimer (highlighted with an arrow) reaches the boundary of $M_1$, and forces all green dimers. The only freedom in dimer placements is the twofold choice on Upper Mystic tiles, and on the boundary. Hence the number of perfect matchings is $2^{N_{\textrm{Mystic}}+1}$.}
\label{fig:inflation}
\end{figure}

The only dimer within either $S_1$ or $M_1$ to meet a boundary vertex appears on $M_1$ (vertex circled in green in Fig.~\ref{fig:implied}). Fig.~\ref{fig:inflation} shows the twice-composition of $S_0$, which we term $S_2$. The special dimer has been highlighted in gold. It forces the two closest green dimers, which in turn force every other green dimer. The result is that all internal edges of $S_2$ not on Upper Mystics are again constrained. 

In fact this behaviour is generic for $S_\mathcal{N}$ regions. Ref.~\onlinecite{spectres} lists all possible ways in which $S_1$ and $M_1$ can meet. The boundaries of $S_1$ and $M_1$ consist only of bivalent or trivalent vertices. The bottom vertex of the Mystic (the rightmost of the two lowest vertices of $M_0$ in Fig.~\ref{fig:implied}) meets the boundary of $S_\mathcal{N}$ exactly once. In all other cases it appears internally. It does so at a trivalent vertex connecting three regions and touches the (gold) boundary dimer of $M_1$. Since a dimer meeting a trivalent vertex forces the absence of dimers on both other edges, the existence of even a single forced dimer along the network of $S_1$ and $M_1$ boundaries is enough to force all remaining dimer placements. The only exception is a twofold freedom along the boundary of the entire connected tile set $S_\mathcal{N}$ (only relevant for finite tile patches). Hence, the total number of dimer matchings is
\begin{align}
\label{eq:count}
\mathcal{Z}_N\left[1\right]=2^{N_{\textrm{Mystic}}+1}
\end{align}
where $N_{\textrm{Mystic}}$ is the number of Mystic tiles $M_0$ (equal to the number of Upper Mystic tiles $M_0^+$).

\begin{figure}[h]
\includegraphics[width=\columnwidth]{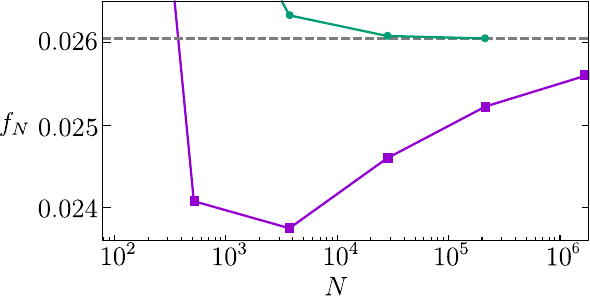}
\caption{The free energy per dimer $f_N$ for patches of spectres containing $N$ vertices. Square points represent compositions $S_2$ to $S_6$. The dashed line shows the analytical result of Eq.~\ref{eq:f}, valid for $N\rightarrow\infty$. Convergence is slow owing to the fractal boundary of $S_{\mathcal{N}\rightarrow\infty}$; green circles represent a series acceleration (see Appendix B). 
}
\label{fig:data}
\end{figure}

In the thermodynamic limit $S_{\mathcal{N}\rightarrow\infty}$, for which the number of vertices $N\rightarrow\infty$, the free energy per dimer is
\begin{align}
\label{eq:f}
f_{\lim N\rightarrow\infty}\left[1\right]=\frac{\textrm{ln}\left(2\right)}{3\left(5+\sqrt{15}\right)}\approx 0.02604
\end{align}
(see Appendix A). To confirm this result we exactly calculated the free energy per dimer numerically in finite patches $S_2$ to $S_6$ using the FKT algorithm~\cite{Kasteleyn61,FisherTemperley61}. The results, shown in Fig.~\ref{fig:data}, converge towards the analytical result. The convergence is slow owing to the fractal boundary of $S_{\mathcal{N}\rightarrow\infty}$, so we also show the result of a series acceleration method~\cite{Aitken26} (Appendix B) which gives a rapid convergence.

The free energy per dimer in the spectre tiling, Eq.~\eqref{eq:f}, is significantly smaller than values obtained in periodic lattices~\cite{Wu06}, e.g.~the square (0.583), honeycomb (0.323), triangular (0.857), and Kagom\'{e} (0.462) lattices. This fits with the observation that all bulk dimers, other than those on $M_0^+$, are completely constrained. 

%
%

\emph{Results (quantum)} --- The QDM can be defined on $S_\mathcal{N}$ by replacing the square tiles (plaquettes) of Ref.~\onlinecite{RokhsarKivelson88} with $S_0$ tiles. Explicitly, on any spectre $S_{0,i}$ we can define $|r_i\rangle$ and $|b_i\rangle$ to be the quantum states with the red and blue dimer placements in Fig.~\ref{fig:implied} respectively. The Hamiltonian then reads
\begin{align}
\label{eq:QDM}
\!\!\!\!\!\!\hat{H}\!\!=\!\!\!\!\!\sum_{S_{0,i}\in S_\mathcal{N}}\!\!\!\!\!-t\left(|r_i\rangle \langle b_i| + |b_i\rangle \langle r_i| \right)
+ V\left(|r_i\rangle \langle r_i|+|b_i\rangle \langle b_i|\right)
\end{align}
where $t$ and $V$ are real and $t$ is positive. The terms weighted by $-t$ can be thought of as defining a kinetic energy operator which enacts `flips' $|r\rangle\leftrightarrow|b\rangle$. The terms weighted by $V$ define a potential energy operator which counts `flippable' plaquettes of the form $|r\rangle$ or $|b\rangle$. 

Eq.~\eqref{eq:QDM} is well studied in the square lattice, where $|r_i\rangle$ denotes two vertical dimers, and $|b_i\rangle$ two horizontal dimers, on square $i$. The so-called Rokhsar Kivelson (RK) point $t=V$ separates ordered phases with different symmetries. The order is set by the sign of $V/t$ which either attempts to maximise or minimise the number of flippable plaquettes~\cite{RokhsarKivelson88,MoessnerRaman07,OakesEA18}. In contrast, on the spectre tiling $S_\mathcal{N}$ the only flippable plaquettes are $M_0^+$ tiles. Their number is entirely fixed. This heavy constraint decouples the problem into one of matching independent $M_0^+$ tiles with quantum dimers. Each tile $M_{0,i}^+$ admits two energy eigenstates which we denote:
\begin{align}
|\pm_i\rangle = \left(|r_i\rangle\pm|b_i\rangle\right)/\sqrt{2}
\end{align}
with corresponding energies $V\mp t$. The ground state of Eq.~\eqref{eq:QDM} is therefore 
\begin{align}
\label{eq:TISE}
\hat{H}\prod_{M_{0,i}^+}|+_i\rangle= (V-t)N_{\textrm{Mystic}}\prod_{M_{0,i}^+}|+_i\rangle.
\end{align}
All excited states can be formed by swapping individual $|+_i\rangle$ for $|-_i\rangle$ at a cost of $2t$ energy per swap.

%
%
%

\emph{Discussion} --- Removing a dimer from a perfect matching results in two unmatched vertices. These can be thought of as particle-like `monomer' defects which can move independently of one another by dimer re-arrangements~\cite{MoessnerRaman07}. Specifically, each monomer lies at the end of an `alternating path', a set of edges alternately uncovered and covered by dimers. Switching which edges are covered and uncovered moves the monomer along the path. The spectre again has an interesting structure in this regard. Note for example that each green alternating path in Fig.~\ref{fig:inflation} terminates only on the boundary and the gold dimer. The same structure holds for all $S_\mathcal{N}$. Deleting the gold dimer to create a pair of monomers, one of the pair can move to the boundary along any green path; in the thermodynamic limit it can escape to infinity. In fact any test pair of monomers has this same feature: one will esape to the boundary, and the other localises on an Upper Mystic $M_0^+$. 

In QDMs on previously studied planar bipartite graphs, such as the square lattice, the RK point $t=V$ constitutes a deconfined quantum critical point between ordered phases~\cite{SenthilEA04}. Deconfinement means that test-monomers can be separated to infinite distance at finite energy cost~\cite{FradkinEA04,BatistaTrugman04,MoessnerRaman07}. In general, since QDMs on bipartite graphs map to compact (matter-free) quantum electrodynamics~\cite{Fradkin,FradkinEA04,SavaryBalents17}, and since deconfined phases cannot exist in compact 2+1D $U(1)$ gauge theories~\cite{Polyakov1977}, the RK point cannot be part of a deconfined phase existing over a range of $V/t$. 

Remarkably, the behaviour in the (bipartite) spectre tiling appears to be at odds with this statement. By the argument just given, any pair of test monomers can be infinitely separated. Doing so preserves the number of flippable plaquettes, so costs no energy according to Eq.~\eqref{eq:QDM} at any $V/t$. Test monomers are therefore deconfined over all $V/t$. We suggest that the result of Ref.~\onlinecite{Polyakov1977} may survive because there seems to be no obvious  mapping to a compact $U(1)$ gauge theory, since the vertices in the spectre tiling connect to variable numbers of edges. Another difference to previous studies is that all previously known bipartite RK points were characterised by algebraic dimer correlations~\cite{MoessnerRaman07}; spectre dimer correlations, being completely uncorrelated between different $M_0^+$ tiles, are not algebraic at any $V/t$. 

QDMs on non-bipartite graphs behave qualitatively differently. Here, they \emph{do} admit deconfined phases spanning a continuous range of $V/t$~\cite{MoessnerSondhi01,MoessnerRaman07,SavaryBalents17,ZhengEA22}, and their emergent gauge field descriptions are $\mathbb{Z}_2$ rather than $U(1)$~\cite{Sachdev92,Misguich02,SavaryBalents17}. The spectre tiling can be made non-bipartite by omitting the gold vertex in Fig.~\ref{fig:implied} whenever it is not forced by the tiling. Fig.~\ref{fig:implied} shows that gold vertices appear only on the boundaries connecting $S_1$ and $M_1$ tiles, so some of the intuition developed here may hold in non-bipartite spectre tilings. Nevertheless, preliminary checks suggest a more complicated behaviour.

\begin{figure}[h]
\includegraphics[width=\columnwidth]{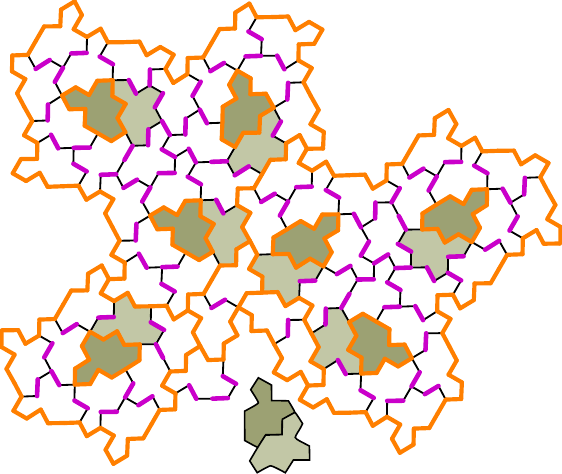}
\caption{Removing the boundary-touching Mystic $M_0$ from region $S_2$ allows the freedom formerly localised to the boundary to move into the bulk. Purple edges always receive a dimer, and black edges never receive a dimer, in any perfect matching. Orange edges are free to either host a dimer or not. 
}
\label{fig:X3}
\end{figure}

Returning to the classical model, different weights $w$ can be assigned in Eq.~\eqref{eq:Z}. For example, in the square lattice different weights might be assigned to horizontal edges compared to vertical edges~\cite{Kasteleyn61}. However, since aperiodic tilings lack a unit cell, there is no obvious choice for assigning weights. One option for $S_\mathcal{N}$ tilings is to assign different weights to edges within regions $S_1$ and $M_1$ while assigning weights consistently between different $S_1$ and $M_1$. Since dimer placements are fixed for all internal edges other than $M_0^+$, the corresponding weights factor out of the partition function. Those edges which never receive a dimer make no contribution regardless of weight. The partition function can therefore also be calculated in this more general case, with the sum being over weights of edges appearing on $M_0^+$ or the boundary of the tiling.

It is interesting to consider what happens when tiles are added or removed from the $S_\mathcal{N}$ regions while still obeying the spectre tiling rules. The total number of vertices can become odd, as in region $M_1$, or even but with an imbalance between the numbers of vertices in the bipartite subgraphs. In both cases there are zero perfect matchings, since dimers connect vertices on distinct bipartite subgraphs. Another possibility is a monomer-free region as shown in Fig.~\ref{fig:X3}. Precisely one $M_0$ touches the boundary of any $S_\mathcal{N}$ region. Removing this $M_0$ from $S_2$, as shown, causes boundaries of some internal $S_1$ and $M_1$ regions to gain a degree of freedom (orange edges host 0 or 1 dimers). This region hosts six perfect matchings excluding those localised around $M_0^+$. In general there is a twofold degree of freedom around any graph cycle (closed loop of edges) which connects to the rest of the graph only via edges not hosting a dimer. This accounts for the freedom around $M_0^+$, the boundaries of $S_\mathcal{N}$, and also these more complicated branching structures in other tile patches. 

The complexity of the spectre tiling leads to a number of surprising simplifications in physical models, permitting exact results where periodic (and other aperiodic) tilings do not. It remains to be seen if there is anything deeper about the structure of the tiling which leads to this simplicity.

%
\textit{Acknowledgments}---The authors thank S.~Franca and J.~Schirmann for helpful discussions, and A.~G.~Grushin, R.~Moessner, P.~d'Ornellas, M.~A.~S\'{a}nchez Mart\'{i}nez, Z.~Ringel, and J.~van Wezel for helpful comments on the manuscript. F.F.~was supported by EPSRC grant EP/X012239/1.

\bibliography{references}

\begin{thebibliography}{36}%
\makeatletter
\providecommand \@ifxundefined [1]{%
 \@ifx{#1\undefined}
}%
\providecommand \@ifnum [1]{%
 \ifnum #1\expandafter \@firstoftwo
 \else \expandafter \@secondoftwo
 \fi
}%
\providecommand \@ifx [1]{%
 \ifx #1\expandafter \@firstoftwo
 \else \expandafter \@secondoftwo
 \fi
}%
\providecommand \natexlab [1]{#1}%
\providecommand \enquote  [1]{``#1''}%
\providecommand \bibnamefont  [1]{#1}%
\providecommand \bibfnamefont [1]{#1}%
\providecommand \citenamefont [1]{#1}%
\providecommand \href@noop [0]{\@secondoftwo}%
\providecommand \href [0]{\begingroup \@sanitize@url \@href}%
\providecommand \@href[1]{\@@startlink{#1}\@@href}%
\providecommand \@@href[1]{\endgroup#1\@@endlink}%
\providecommand \@sanitize@url [0]{\catcode `\\12\catcode `\$12\catcode `\&12\catcode `\#12\catcode `\^12\catcode `\_12\catcode `\%12\relax}%
\providecommand \@@startlink[1]{}%
\providecommand \@@endlink[0]{}%
\providecommand \url  [0]{\begingroup\@sanitize@url \@url }%
\providecommand \@url [1]{\endgroup\@href {#1}{\urlprefix }}%
\providecommand \urlprefix  [0]{URL }%
\providecommand \Eprint [0]{\href }%
\providecommand \doibase [0]{https://doi.org/}%
\providecommand \selectlanguage [0]{\@gobble}%
\providecommand \bibinfo  [0]{\@secondoftwo}%
\providecommand \bibfield  [0]{\@secondoftwo}%
\providecommand \translation [1]{[#1]}%
\providecommand \BibitemOpen [0]{}%
\providecommand \bibitemStop [0]{}%
\providecommand \bibitemNoStop [0]{.\EOS\space}%
\providecommand \EOS [0]{\spacefactor3000\relax}%
\providecommand \BibitemShut  [1]{\csname bibitem#1\endcsname}%
\let\auto@bib@innerbib\@empty
\bibitem [{\citenamefont {Kasteleyn}(1961)}]{Kasteleyn61}%
  \BibitemOpen
  \bibfield  {author} {\bibinfo {author} {\bibfnamefont {P.}~\bibnamefont {Kasteleyn}},\ }\bibfield  {title} {\bibinfo {title} {{The statistics of dimers on a lattice: I. The number of dimer arrangements on a quadratic lattice}},\ }\href {https://doi.org/https://doi.org/10.1016/0031-8914(61)90063-5} {\bibfield  {journal} {\bibinfo  {journal} {Physica}\ }\textbf {\bibinfo {volume} {27}},\ \bibinfo {pages} {1209 } (\bibinfo {year} {1961})}\BibitemShut {NoStop}%
\bibitem [{\citenamefont {Fisher}(1961)}]{Fisher61}%
  \BibitemOpen
  \bibfield  {author} {\bibinfo {author} {\bibfnamefont {M.~E.}\ \bibnamefont {Fisher}},\ }\bibfield  {title} {\bibinfo {title} {{Statistical Mechanics of Dimers on a Plane Lattice}},\ }\href {https://doi.org/10.1103/PhysRev.124.1664} {\bibfield  {journal} {\bibinfo  {journal} {Phys. Rev.}\ }\textbf {\bibinfo {volume} {124}},\ \bibinfo {pages} {1664} (\bibinfo {year} {1961})}\BibitemShut {NoStop}%
\bibitem [{\citenamefont {Temperley}\ and\ \citenamefont {Fisher}(1961)}]{FisherTemperley61}%
  \BibitemOpen
  \bibfield  {author} {\bibinfo {author} {\bibfnamefont {H.~N.~V.}\ \bibnamefont {Temperley}}\ and\ \bibinfo {author} {\bibfnamefont {M.~E.}\ \bibnamefont {Fisher}},\ }\bibfield  {title} {\bibinfo {title} {{Dimer problem in statistical mechanics - an exact result}},\ }\href {https://doi.org/10.1080/14786436108243366} {\bibfield  {journal} {\bibinfo  {journal} {The Philosophical Magazine: A Journal of Theoretical Experimental and Applied Physics}\ }\textbf {\bibinfo {volume} {6}},\ \bibinfo {pages} {1061} (\bibinfo {year} {1961})}\BibitemShut {NoStop}%
\bibitem [{\citenamefont {Kasteleyn}(1963)}]{Kasteleyn63}%
  \BibitemOpen
  \bibfield  {author} {\bibinfo {author} {\bibfnamefont {P.~W.}\ \bibnamefont {Kasteleyn}},\ }\bibfield  {title} {\bibinfo {title} {{Dimer Statistics and Phase Transitions}},\ }\href {https://doi.org/10.1063/1.1703953} {\bibfield  {journal} {\bibinfo  {journal} {Journal of Mathematical Physics}\ }\textbf {\bibinfo {volume} {4}},\ \bibinfo {pages} {287} (\bibinfo {year} {1963})},\ \Eprint {https://arxiv.org/abs/https://doi.org/10.1063/1.1703953} {https://doi.org/10.1063/1.1703953} \BibitemShut {NoStop}%
\bibitem [{\citenamefont {Kasteleyn}(1967)}]{Kasteleyn67}%
  \BibitemOpen
  \bibfield  {author} {\bibinfo {author} {\bibfnamefont {P.}~\bibnamefont {Kasteleyn}},\ }\bibfield  {title} {\bibinfo {title} {{Graph theory and crystal physics}},\ }\href@noop {} {\bibfield  {journal} {\bibinfo  {journal} {Graph Theory and Theoretical Physics}\ ,\ \bibinfo {pages} {43}} (\bibinfo {year} {1967})}\BibitemShut {NoStop}%
\bibitem [{\citenamefont {Longuet-Higgins}(1950)}]{LH50}%
  \BibitemOpen
  \bibfield  {author} {\bibinfo {author} {\bibfnamefont {H.~C.}\ \bibnamefont {Longuet-Higgins}},\ }\bibfield  {title} {\bibinfo {title} {{Some studies in molecular orbital theory. I. Resonance structures and molecular orbitals in un-saturated hydrocarbons}},\ }\href@noop {} {\bibfield  {journal} {\bibinfo  {journal} {J. Chem. Phys.}\ }\textbf {\bibinfo {volume} {18}},\ \bibinfo {pages} {265} (\bibinfo {year} {1950})}\BibitemShut {NoStop}%
\bibitem [{\citenamefont {Lieb}(1989)}]{Lieb89}%
  \BibitemOpen
  \bibfield  {author} {\bibinfo {author} {\bibfnamefont {E.~H.}\ \bibnamefont {Lieb}},\ }\bibfield  {title} {\bibinfo {title} {{Two theorems on the Hubbard model}},\ }\href {https://doi.org/10.1103/PhysRevLett.62.1201} {\bibfield  {journal} {\bibinfo  {journal} {Phys. Rev. Lett.}\ }\textbf {\bibinfo {volume} {62}},\ \bibinfo {pages} {1201} (\bibinfo {year} {1989})}\BibitemShut {NoStop}%
\bibitem [{\citenamefont {Schirmann}\ \emph {et~al.}(2023)\citenamefont {Schirmann}, \citenamefont {Franca}, \citenamefont {Flicker},\ and\ \citenamefont {Grushin}}]{SchirmannEA23}%
  \BibitemOpen
  \bibfield  {author} {\bibinfo {author} {\bibfnamefont {J.}~\bibnamefont {Schirmann}}, \bibinfo {author} {\bibfnamefont {S.}~\bibnamefont {Franca}}, \bibinfo {author} {\bibfnamefont {F.}~\bibnamefont {Flicker}},\ and\ \bibinfo {author} {\bibfnamefont {A.~G.}\ \bibnamefont {Grushin}},\ }\bibfield  {title} {\bibinfo {title} {{Physical properties of the Hat aperiodic monotile: Graphene-like features, chirality and zero-modes}},\ }\href@noop {} {\bibfield  {journal} {\bibinfo  {journal} {arXiv:2307.11054 [cond-mat.mes-hall]}\ } (\bibinfo {year} {2023})}\BibitemShut {NoStop}%
\bibitem [{\citenamefont {Baxter}(1982)}]{Baxter}%
  \BibitemOpen
  \bibfield  {author} {\bibinfo {author} {\bibfnamefont {R.~J.}\ \bibnamefont {Baxter}},\ }\href@noop {} {\emph {\bibinfo {title} {{Exactly Solved Models in Statistical Mechanics}}}}\ (\bibinfo  {publisher} {Academic Press, Harcourt Brace Jovanivich (London)},\ \bibinfo {year} {1982})\BibitemShut {NoStop}%
\bibitem [{\citenamefont {Rokhsar}\ and\ \citenamefont {Kivelson}(1988)}]{RokhsarKivelson88}%
  \BibitemOpen
  \bibfield  {author} {\bibinfo {author} {\bibfnamefont {D.~S.}\ \bibnamefont {Rokhsar}}\ and\ \bibinfo {author} {\bibfnamefont {S.~A.}\ \bibnamefont {Kivelson}},\ }\bibfield  {title} {\bibinfo {title} {{Superconductivity and the Quantum Hard-Core Dimer Gas}},\ }\href {https://doi.org/10.1103/PhysRevLett.61.2376} {\bibfield  {journal} {\bibinfo  {journal} {Phys. Rev. Lett.}\ }\textbf {\bibinfo {volume} {61}},\ \bibinfo {pages} {2376} (\bibinfo {year} {1988})}\BibitemShut {NoStop}%
\bibitem [{\citenamefont {Moessner}\ and\ \citenamefont {Raman}(2007)}]{MoessnerRaman07}%
  \BibitemOpen
  \bibfield  {author} {\bibinfo {author} {\bibfnamefont {R.}~\bibnamefont {Moessner}}\ and\ \bibinfo {author} {\bibfnamefont {K.~S.}\ \bibnamefont {Raman}},\ }\href@noop {} {\emph {\bibinfo {title} {{Quantum Dimer Models}}}}\ (\bibinfo  {publisher} {Lecture Notes, Trieste},\ \bibinfo {year} {2007})\BibitemShut {NoStop}%
\bibitem [{\citenamefont {Anderson}(1987)}]{Anderson87}%
  \BibitemOpen
  \bibfield  {author} {\bibinfo {author} {\bibfnamefont {P.}~\bibnamefont {Anderson}},\ }\bibfield  {title} {\bibinfo {title} {{The Resonating Valence Bond State in La$_2$CuO$_4$ and Superconductivity}},\ }\href@noop {} {\bibfield  {journal} {\bibinfo  {journal} {Science}\ }\textbf {\bibinfo {volume} {235}},\ \bibinfo {pages} {1196} (\bibinfo {year} {1987})}\BibitemShut {NoStop}%
\bibitem [{\citenamefont {Savary}\ and\ \citenamefont {Balents}(2017)}]{SavaryBalents17}%
  \BibitemOpen
  \bibfield  {author} {\bibinfo {author} {\bibfnamefont {L.}~\bibnamefont {Savary}}\ and\ \bibinfo {author} {\bibfnamefont {L.}~\bibnamefont {Balents}},\ }\bibfield  {title} {\bibinfo {title} {{Quantum spin liquids: a review}},\ }\href@noop {} {\bibfield  {journal} {\bibinfo  {journal} {Rep. Prog. Phys.}\ }\textbf {\bibinfo {volume} {80}},\ \bibinfo {pages} {016502} (\bibinfo {year} {2017})}\BibitemShut {NoStop}%
\bibitem [{\citenamefont {Satzinger}\ \emph {et~al.}(2021)\citenamefont {Satzinger}, \citenamefont {Liu}, \citenamefont {Smith}, \citenamefont {Knapp}, \citenamefont {Newman}, \citenamefont {Jones}, \citenamefont {Chen}, \citenamefont {Quintana}, \citenamefont {Mi}, \citenamefont {Dunsworth}, \citenamefont {Gidney}, \citenamefont {Aleiner}, \citenamefont {Arute}, \citenamefont {Arya}, \citenamefont {Atalaya}, \citenamefont {Babbush}, \citenamefont {Bardin}, \citenamefont {Barends}, \citenamefont {Basso}, \citenamefont {Bengtsson}, \citenamefont {Bilmes}, \citenamefont {Broughton}, \citenamefont {Buckley}, \citenamefont {Buell}, \citenamefont {Burkett}, \citenamefont {Bushnell}, \citenamefont {Chiaro}, \citenamefont {Collins}, \citenamefont {Courtney}, \citenamefont {Demura}, \citenamefont {Derk}, \citenamefont {Eppens}, \citenamefont {Erickson}, \citenamefont {Faoro}, \citenamefont {Farhi}, \citenamefont {Fowler}, \citenamefont {Foxen}, \citenamefont {Giustina}, \citenamefont {Greene}, \citenamefont {Gross},
  \citenamefont {Harrigan}, \citenamefont {Harrington}, \citenamefont {Hilton}, \citenamefont {Hong}, \citenamefont {Huang}, \citenamefont {Huggins}, \citenamefont {Ioffe}, \citenamefont {Isakov}, \citenamefont {Jeffrey}, \citenamefont {Jiang}, \citenamefont {Kafri}, \citenamefont {Kechedzhi}, \citenamefont {Khattar}, \citenamefont {Kim}, \citenamefont {Klimov}, \citenamefont {Korotkov}, \citenamefont {Kostritsa}, \citenamefont {Landhuis}, \citenamefont {Laptev}, \citenamefont {Locharla}, \citenamefont {Lucero}, \citenamefont {Martin}, \citenamefont {McClean}, \citenamefont {McEwen}, \citenamefont {Miao}, \citenamefont {Mohseni}, \citenamefont {Montazeri}, \citenamefont {Mruczkiewicz}, \citenamefont {Mutus}, \citenamefont {Naaman}, \citenamefont {Neeley}, \citenamefont {Neill}, \citenamefont {Niu}, \citenamefont {O’Brien}, \citenamefont {Opremcak}, \citenamefont {Pató}, \citenamefont {Petukhov}, \citenamefont {Rubin}, \citenamefont {Sank}, \citenamefont {Shvarts}, \citenamefont {Strain}, \citenamefont
  {Szalay}, \citenamefont {Villalonga}, \citenamefont {White}, \citenamefont {Yao}, \citenamefont {Yeh}, \citenamefont {Yoo}, \citenamefont {Zalcman}, \citenamefont {Neven}, \citenamefont {Boixo}, \citenamefont {Megrant}, \citenamefont {Chen}, \citenamefont {Kelly}, \citenamefont {Smelyanskiy}, \citenamefont {Kitaev}, \citenamefont {Knap}, \citenamefont {Pollmann},\ and\ \citenamefont {Roushan}}]{SatzingerEA21}%
  \BibitemOpen
  \bibfield  {author} {\bibinfo {author} {\bibfnamefont {K.~J.}\ \bibnamefont {Satzinger}}, \bibinfo {author} {\bibfnamefont {Y.-J.}\ \bibnamefont {Liu}}, \bibinfo {author} {\bibfnamefont {A.}~\bibnamefont {Smith}}, \bibinfo {author} {\bibfnamefont {C.}~\bibnamefont {Knapp}}, \bibinfo {author} {\bibfnamefont {M.}~\bibnamefont {Newman}}, \bibinfo {author} {\bibfnamefont {C.}~\bibnamefont {Jones}}, \bibinfo {author} {\bibfnamefont {Z.}~\bibnamefont {Chen}}, \bibinfo {author} {\bibfnamefont {C.}~\bibnamefont {Quintana}}, \bibinfo {author} {\bibfnamefont {X.}~\bibnamefont {Mi}}, \bibinfo {author} {\bibfnamefont {A.}~\bibnamefont {Dunsworth}}, \bibinfo {author} {\bibfnamefont {C.}~\bibnamefont {Gidney}}, \bibinfo {author} {\bibfnamefont {I.}~\bibnamefont {Aleiner}}, \bibinfo {author} {\bibfnamefont {F.}~\bibnamefont {Arute}}, \bibinfo {author} {\bibfnamefont {K.}~\bibnamefont {Arya}}, \bibinfo {author} {\bibfnamefont {J.}~\bibnamefont {Atalaya}}, \bibinfo {author} {\bibfnamefont {R.}~\bibnamefont {Babbush}},
  \bibinfo {author} {\bibfnamefont {J.~C.}\ \bibnamefont {Bardin}}, \bibinfo {author} {\bibfnamefont {R.}~\bibnamefont {Barends}}, \bibinfo {author} {\bibfnamefont {J.}~\bibnamefont {Basso}}, \bibinfo {author} {\bibfnamefont {A.}~\bibnamefont {Bengtsson}}, \bibinfo {author} {\bibfnamefont {A.}~\bibnamefont {Bilmes}}, \bibinfo {author} {\bibfnamefont {M.}~\bibnamefont {Broughton}}, \bibinfo {author} {\bibfnamefont {B.~B.}\ \bibnamefont {Buckley}}, \bibinfo {author} {\bibfnamefont {D.~A.}\ \bibnamefont {Buell}}, \bibinfo {author} {\bibfnamefont {B.}~\bibnamefont {Burkett}}, \bibinfo {author} {\bibfnamefont {N.}~\bibnamefont {Bushnell}}, \bibinfo {author} {\bibfnamefont {B.}~\bibnamefont {Chiaro}}, \bibinfo {author} {\bibfnamefont {R.}~\bibnamefont {Collins}}, \bibinfo {author} {\bibfnamefont {W.}~\bibnamefont {Courtney}}, \bibinfo {author} {\bibfnamefont {S.}~\bibnamefont {Demura}}, \bibinfo {author} {\bibfnamefont {A.~R.}\ \bibnamefont {Derk}}, \bibinfo {author} {\bibfnamefont {D.}~\bibnamefont {Eppens}},
  \bibinfo {author} {\bibfnamefont {C.}~\bibnamefont {Erickson}}, \bibinfo {author} {\bibfnamefont {L.}~\bibnamefont {Faoro}}, \bibinfo {author} {\bibfnamefont {E.}~\bibnamefont {Farhi}}, \bibinfo {author} {\bibfnamefont {A.~G.}\ \bibnamefont {Fowler}}, \bibinfo {author} {\bibfnamefont {B.}~\bibnamefont {Foxen}}, \bibinfo {author} {\bibfnamefont {M.}~\bibnamefont {Giustina}}, \bibinfo {author} {\bibfnamefont {A.}~\bibnamefont {Greene}}, \bibinfo {author} {\bibfnamefont {J.~A.}\ \bibnamefont {Gross}}, \bibinfo {author} {\bibfnamefont {M.~P.}\ \bibnamefont {Harrigan}}, \bibinfo {author} {\bibfnamefont {S.~D.}\ \bibnamefont {Harrington}}, \bibinfo {author} {\bibfnamefont {J.}~\bibnamefont {Hilton}}, \bibinfo {author} {\bibfnamefont {S.}~\bibnamefont {Hong}}, \bibinfo {author} {\bibfnamefont {T.}~\bibnamefont {Huang}}, \bibinfo {author} {\bibfnamefont {W.~J.}\ \bibnamefont {Huggins}}, \bibinfo {author} {\bibfnamefont {L.~B.}\ \bibnamefont {Ioffe}}, \bibinfo {author} {\bibfnamefont {S.~V.}\ \bibnamefont {Isakov}},
  \bibinfo {author} {\bibfnamefont {E.}~\bibnamefont {Jeffrey}}, \bibinfo {author} {\bibfnamefont {Z.}~\bibnamefont {Jiang}}, \bibinfo {author} {\bibfnamefont {D.}~\bibnamefont {Kafri}}, \bibinfo {author} {\bibfnamefont {K.}~\bibnamefont {Kechedzhi}}, \bibinfo {author} {\bibfnamefont {T.}~\bibnamefont {Khattar}}, \bibinfo {author} {\bibfnamefont {S.}~\bibnamefont {Kim}}, \bibinfo {author} {\bibfnamefont {P.~V.}\ \bibnamefont {Klimov}}, \bibinfo {author} {\bibfnamefont {A.~N.}\ \bibnamefont {Korotkov}}, \bibinfo {author} {\bibfnamefont {F.}~\bibnamefont {Kostritsa}}, \bibinfo {author} {\bibfnamefont {D.}~\bibnamefont {Landhuis}}, \bibinfo {author} {\bibfnamefont {P.}~\bibnamefont {Laptev}}, \bibinfo {author} {\bibfnamefont {A.}~\bibnamefont {Locharla}}, \bibinfo {author} {\bibfnamefont {E.}~\bibnamefont {Lucero}}, \bibinfo {author} {\bibfnamefont {O.}~\bibnamefont {Martin}}, \bibinfo {author} {\bibfnamefont {J.~R.}\ \bibnamefont {McClean}}, \bibinfo {author} {\bibfnamefont {M.}~\bibnamefont {McEwen}}, \bibinfo
  {author} {\bibfnamefont {K.~C.}\ \bibnamefont {Miao}}, \bibinfo {author} {\bibfnamefont {M.}~\bibnamefont {Mohseni}}, \bibinfo {author} {\bibfnamefont {S.}~\bibnamefont {Montazeri}}, \bibinfo {author} {\bibfnamefont {W.}~\bibnamefont {Mruczkiewicz}}, \bibinfo {author} {\bibfnamefont {J.}~\bibnamefont {Mutus}}, \bibinfo {author} {\bibfnamefont {O.}~\bibnamefont {Naaman}}, \bibinfo {author} {\bibfnamefont {M.}~\bibnamefont {Neeley}}, \bibinfo {author} {\bibfnamefont {C.}~\bibnamefont {Neill}}, \bibinfo {author} {\bibfnamefont {M.~Y.}\ \bibnamefont {Niu}}, \bibinfo {author} {\bibfnamefont {T.~E.}\ \bibnamefont {O’Brien}}, \bibinfo {author} {\bibfnamefont {A.}~\bibnamefont {Opremcak}}, \bibinfo {author} {\bibfnamefont {B.}~\bibnamefont {Pató}}, \bibinfo {author} {\bibfnamefont {A.}~\bibnamefont {Petukhov}}, \bibinfo {author} {\bibfnamefont {N.~C.}\ \bibnamefont {Rubin}}, \bibinfo {author} {\bibfnamefont {D.}~\bibnamefont {Sank}}, \bibinfo {author} {\bibfnamefont {V.}~\bibnamefont {Shvarts}}, \bibinfo
  {author} {\bibfnamefont {D.}~\bibnamefont {Strain}}, \bibinfo {author} {\bibfnamefont {M.}~\bibnamefont {Szalay}}, \bibinfo {author} {\bibfnamefont {B.}~\bibnamefont {Villalonga}}, \bibinfo {author} {\bibfnamefont {T.~C.}\ \bibnamefont {White}}, \bibinfo {author} {\bibfnamefont {Z.}~\bibnamefont {Yao}}, \bibinfo {author} {\bibfnamefont {P.}~\bibnamefont {Yeh}}, \bibinfo {author} {\bibfnamefont {J.}~\bibnamefont {Yoo}}, \bibinfo {author} {\bibfnamefont {A.}~\bibnamefont {Zalcman}}, \bibinfo {author} {\bibfnamefont {H.}~\bibnamefont {Neven}}, \bibinfo {author} {\bibfnamefont {S.}~\bibnamefont {Boixo}}, \bibinfo {author} {\bibfnamefont {A.}~\bibnamefont {Megrant}}, \bibinfo {author} {\bibfnamefont {Y.}~\bibnamefont {Chen}}, \bibinfo {author} {\bibfnamefont {J.}~\bibnamefont {Kelly}}, \bibinfo {author} {\bibfnamefont {V.}~\bibnamefont {Smelyanskiy}}, \bibinfo {author} {\bibfnamefont {A.}~\bibnamefont {Kitaev}}, \bibinfo {author} {\bibfnamefont {M.}~\bibnamefont {Knap}}, \bibinfo {author} {\bibfnamefont
  {F.}~\bibnamefont {Pollmann}},\ and\ \bibinfo {author} {\bibfnamefont {P.}~\bibnamefont {Roushan}},\ }\bibfield  {title} {\bibinfo {title} {Realizing topologically ordered states on a quantum processor},\ }\href {https://doi.org/10.1126/science.abi8378} {\bibfield  {journal} {\bibinfo  {journal} {Science}\ }\textbf {\bibinfo {volume} {374}},\ \bibinfo {pages} {1237} (\bibinfo {year} {2021})},\ \Eprint {https://arxiv.org/abs/https://www.science.org/doi/pdf/10.1126/science.abi8378} {https://www.science.org/doi/pdf/10.1126/science.abi8378} \BibitemShut {NoStop}%
\bibitem [{\citenamefont {Semeghini}\ \emph {et~al.}(2021)\citenamefont {Semeghini}, \citenamefont {Levine}, \citenamefont {Keesling}, \citenamefont {Ebadi}, \citenamefont {Wang}, \citenamefont {Bluvstein}, \citenamefont {Verresen}, \citenamefont {Pichler}, \citenamefont {Kalinowski}, \citenamefont {Samajdar}, \citenamefont {Omran}, \citenamefont {Sachdev}, \citenamefont {Vishwanath}, \citenamefont {Greiner}, \citenamefont {Vuletić},\ and\ \citenamefont {Lukin}}]{SemeghiniEA21}%
  \BibitemOpen
  \bibfield  {author} {\bibinfo {author} {\bibfnamefont {G.}~\bibnamefont {Semeghini}}, \bibinfo {author} {\bibfnamefont {H.}~\bibnamefont {Levine}}, \bibinfo {author} {\bibfnamefont {A.}~\bibnamefont {Keesling}}, \bibinfo {author} {\bibfnamefont {S.}~\bibnamefont {Ebadi}}, \bibinfo {author} {\bibfnamefont {T.~T.}\ \bibnamefont {Wang}}, \bibinfo {author} {\bibfnamefont {D.}~\bibnamefont {Bluvstein}}, \bibinfo {author} {\bibfnamefont {R.}~\bibnamefont {Verresen}}, \bibinfo {author} {\bibfnamefont {H.}~\bibnamefont {Pichler}}, \bibinfo {author} {\bibfnamefont {M.}~\bibnamefont {Kalinowski}}, \bibinfo {author} {\bibfnamefont {R.}~\bibnamefont {Samajdar}}, \bibinfo {author} {\bibfnamefont {A.}~\bibnamefont {Omran}}, \bibinfo {author} {\bibfnamefont {S.}~\bibnamefont {Sachdev}}, \bibinfo {author} {\bibfnamefont {A.}~\bibnamefont {Vishwanath}}, \bibinfo {author} {\bibfnamefont {M.}~\bibnamefont {Greiner}}, \bibinfo {author} {\bibfnamefont {V.}~\bibnamefont {Vuletić}},\ and\ \bibinfo {author} {\bibfnamefont
  {M.~D.}\ \bibnamefont {Lukin}},\ }\bibfield  {title} {\bibinfo {title} {{Probing topological spin liquids on a programmable quantum simulator}},\ }\href {https://doi.org/10.1126/science.abi8794} {\bibfield  {journal} {\bibinfo  {journal} {Science}\ }\textbf {\bibinfo {volume} {374}},\ \bibinfo {pages} {1242} (\bibinfo {year} {2021})},\ \Eprint {https://arxiv.org/abs/https://www.science.org/doi/pdf/10.1126/science.abi8794} {https://www.science.org/doi/pdf/10.1126/science.abi8794} \BibitemShut {NoStop}%
\bibitem [{\citenamefont {Wu}(2006)}]{Wu06}%
  \BibitemOpen
  \bibfield  {author} {\bibinfo {author} {\bibfnamefont {F.~Y.}\ \bibnamefont {Wu}},\ }\bibfield  {title} {\bibinfo {title} {{Dimers on Two-Dimensional Lattices}},\ }\href@noop {} {\bibfield  {journal} {\bibinfo  {journal} {International Journal of Modern Physics B}\ }\textbf {\bibinfo {volume} {20}},\ \bibinfo {pages} {5357} (\bibinfo {year} {2006})}\BibitemShut {NoStop}%
\bibitem [{\citenamefont {Flicker}\ \emph {et~al.}(2020)\citenamefont {Flicker}, \citenamefont {Simon},\ and\ \citenamefont {Parameswaran}}]{FlickerEA20}%
  \BibitemOpen
  \bibfield  {author} {\bibinfo {author} {\bibfnamefont {F.}~\bibnamefont {Flicker}}, \bibinfo {author} {\bibfnamefont {S.~H.}\ \bibnamefont {Simon}},\ and\ \bibinfo {author} {\bibfnamefont {S.}~\bibnamefont {Parameswaran}},\ }\bibfield  {title} {\bibinfo {title} {{Classical dimers on penrose tilings}},\ }\href@noop {} {\bibfield  {journal} {\bibinfo  {journal} {Physical Review X}\ }\textbf {\bibinfo {volume} {10}},\ \bibinfo {pages} {011005} (\bibinfo {year} {2020})}\BibitemShut {NoStop}%
\bibitem [{\citenamefont {Lloyd}\ \emph {et~al.}(2022)\citenamefont {Lloyd}, \citenamefont {Biswas}, \citenamefont {Simon}, \citenamefont {Parameswaran},\ and\ \citenamefont {Flicker}}]{LloydEA21}%
  \BibitemOpen
  \bibfield  {author} {\bibinfo {author} {\bibfnamefont {J.}~\bibnamefont {Lloyd}}, \bibinfo {author} {\bibfnamefont {S.}~\bibnamefont {Biswas}}, \bibinfo {author} {\bibfnamefont {S.~H.}\ \bibnamefont {Simon}}, \bibinfo {author} {\bibfnamefont {S.~A.}\ \bibnamefont {Parameswaran}},\ and\ \bibinfo {author} {\bibfnamefont {F.}~\bibnamefont {Flicker}},\ }\bibfield  {title} {\bibinfo {title} {{Statistical mechanics of dimers on quasiperiodic Ammann-Beenker tilings}},\ }\href {https://doi.org/10.1103/PhysRevB.106.094202} {\bibfield  {journal} {\bibinfo  {journal} {Phys. Rev. B}\ }\textbf {\bibinfo {volume} {106}},\ \bibinfo {pages} {094202} (\bibinfo {year} {2022})}\BibitemShut {NoStop}%
\bibitem [{\citenamefont {Singh}\ \emph {et~al.}(2023)\citenamefont {Singh}, \citenamefont {Lloyd},\ and\ \citenamefont {Flicker}}]{SinghEA23}%
  \BibitemOpen
  \bibfield  {author} {\bibinfo {author} {\bibfnamefont {S.}~\bibnamefont {Singh}}, \bibinfo {author} {\bibfnamefont {J.}~\bibnamefont {Lloyd}},\ and\ \bibinfo {author} {\bibfnamefont {F.}~\bibnamefont {Flicker}},\ }\bibfield  {title} {\bibinfo {title} {{Hamiltonian Cycles on Ammann Beenker Tilings}},\ }\href@noop {} {\bibfield  {journal} {\bibinfo  {journal} {arXiv:2302.01940 [cond-mat.stat-mech]}\ } (\bibinfo {year} {2023})}\BibitemShut {NoStop}%
\bibitem [{\citenamefont {Baake}\ and\ \citenamefont {Grimm}(2013)}]{BaakeGrimm}%
  \BibitemOpen
  \bibfield  {author} {\bibinfo {author} {\bibfnamefont {M.}~\bibnamefont {Baake}}\ and\ \bibinfo {author} {\bibfnamefont {U.}~\bibnamefont {Grimm}},\ }\href@noop {} {\emph {\bibinfo {title} {{Aperiodic Order Volume 1: A Mathematical Invitation}}}}\ (\bibinfo  {publisher} {Cambridge University Press, Cambridge},\ \bibinfo {year} {2013})\BibitemShut {NoStop}%
\bibitem [{\citenamefont {Senechal}(1995)}]{Senechal}%
  \BibitemOpen
  \bibfield  {author} {\bibinfo {author} {\bibfnamefont {M.}~\bibnamefont {Senechal}},\ }\href@noop {} {\emph {\bibinfo {title} {{Quasicrystals and Geometry}}}}\ (\bibinfo  {publisher} {{Cambridge University Press}},\ \bibinfo {year} {1995})\BibitemShut {NoStop}%
\bibitem [{\citenamefont {Penrose}(1974)}]{Penrose74}%
  \BibitemOpen
  \bibfield  {author} {\bibinfo {author} {\bibfnamefont {R.}~\bibnamefont {Penrose}},\ }\bibfield  {title} {\bibinfo {title} {{The role of aesthetics in pure and applied mathematical research}},\ }\href@noop {} {\bibfield  {journal} {\bibinfo  {journal} {Bulletin of the Institute of Mathematics and its Applications}\ }\textbf {\bibinfo {volume} {10}},\ \bibinfo {pages} {266ff} (\bibinfo {year} {1974})}\BibitemShut {NoStop}%
\bibitem [{\citenamefont {Gr\"{u}nbaum}\ and\ \citenamefont {Shephard}(1986)}]{GrunbaumShephard}%
  \BibitemOpen
  \bibfield  {author} {\bibinfo {author} {\bibfnamefont {B.}~\bibnamefont {Gr\"{u}nbaum}}\ and\ \bibinfo {author} {\bibfnamefont {G.~C.}\ \bibnamefont {Shephard}},\ }\href@noop {} {\emph {\bibinfo {title} {{Tilings and Patterns}}}}\ (\bibinfo  {publisher} {W. H. Freeman and Company, New York},\ \bibinfo {year} {1986})\BibitemShut {NoStop}%
\bibitem [{\citenamefont {Smith}\ \emph {et~al.}(2023{\natexlab{a}})\citenamefont {Smith}, \citenamefont {Myers}, \citenamefont {Kaplan},\ and\ \citenamefont {Goodman-Strauss}}]{spectres}%
  \BibitemOpen
  \bibfield  {author} {\bibinfo {author} {\bibfnamefont {D.}~\bibnamefont {Smith}}, \bibinfo {author} {\bibfnamefont {J.~S.}\ \bibnamefont {Myers}}, \bibinfo {author} {\bibfnamefont {C.~S.}\ \bibnamefont {Kaplan}},\ and\ \bibinfo {author} {\bibfnamefont {C.}~\bibnamefont {Goodman-Strauss}},\ }\bibfield  {title} {\bibinfo {title} {{A chiral aperiodic monotile}},\ }\href@noop {} {\bibfield  {journal} {\bibinfo  {journal} {arXiv:2305.17743 [math.CO]}\ } (\bibinfo {year} {2023}{\natexlab{a}})}\BibitemShut {NoStop}%
\bibitem [{\citenamefont {Smith}\ \emph {et~al.}(2023{\natexlab{b}})\citenamefont {Smith}, \citenamefont {Myers}, \citenamefont {Kaplan},\ and\ \citenamefont {Goodman-Strauss}}]{hats}%
  \BibitemOpen
  \bibfield  {author} {\bibinfo {author} {\bibfnamefont {D.}~\bibnamefont {Smith}}, \bibinfo {author} {\bibfnamefont {J.~S.}\ \bibnamefont {Myers}}, \bibinfo {author} {\bibfnamefont {C.~S.}\ \bibnamefont {Kaplan}},\ and\ \bibinfo {author} {\bibfnamefont {C.}~\bibnamefont {Goodman-Strauss}},\ }\bibfield  {title} {\bibinfo {title} {{An aperiodic monotile}},\ }\href@noop {} {\bibfield  {journal} {\bibinfo  {journal} {arXiv:2303.10798 [math.CO]}\ } (\bibinfo {year} {2023}{\natexlab{b}})}\BibitemShut {NoStop}%
\bibitem [{\citenamefont {Aitken}(1926)}]{Aitken26}%
  \BibitemOpen
  \bibfield  {author} {\bibinfo {author} {\bibfnamefont {A.}~\bibnamefont {Aitken}},\ }\bibfield  {title} {\bibinfo {title} {{On Bernoulli's numerical solution of algebraic equations}},\ }\href {https://doi.org/10.1017/S0370164600022070} {\bibfield  {journal} {\bibinfo  {journal} {Proceedings of the Royal Society of Edinburgh}\ }\textbf {\bibinfo {volume} {46}},\ \bibinfo {pages} {289} (\bibinfo {year} {1926})}\BibitemShut {NoStop}%
\bibitem [{\citenamefont {Oakes}\ \emph {et~al.}(2018)\citenamefont {Oakes}, \citenamefont {Powell}, \citenamefont {Castelnovo}, \citenamefont {Lamacraft},\ and\ \citenamefont {Garrahan}}]{OakesEA18}%
  \BibitemOpen
  \bibfield  {author} {\bibinfo {author} {\bibfnamefont {T.}~\bibnamefont {Oakes}}, \bibinfo {author} {\bibfnamefont {S.}~\bibnamefont {Powell}}, \bibinfo {author} {\bibfnamefont {C.}~\bibnamefont {Castelnovo}}, \bibinfo {author} {\bibfnamefont {A.}~\bibnamefont {Lamacraft}},\ and\ \bibinfo {author} {\bibfnamefont {J.~P.}\ \bibnamefont {Garrahan}},\ }\bibfield  {title} {\bibinfo {title} {{Phases of quantum dimers from ensembles of classical stochastic trajectories}},\ }\href {https://doi.org/10.1103/PhysRevB.98.064302} {\bibfield  {journal} {\bibinfo  {journal} {Phys. Rev. B}\ }\textbf {\bibinfo {volume} {98}},\ \bibinfo {pages} {064302} (\bibinfo {year} {2018})}\BibitemShut {NoStop}%
\bibitem [{\citenamefont {Senthil}\ \emph {et~al.}(2004)\citenamefont {Senthil}, \citenamefont {Balents}, \citenamefont {Sachdev}, \citenamefont {Vishwanath},\ and\ \citenamefont {Fisher}}]{SenthilEA04}%
  \BibitemOpen
  \bibfield  {author} {\bibinfo {author} {\bibfnamefont {T.}~\bibnamefont {Senthil}}, \bibinfo {author} {\bibfnamefont {L.}~\bibnamefont {Balents}}, \bibinfo {author} {\bibfnamefont {S.}~\bibnamefont {Sachdev}}, \bibinfo {author} {\bibfnamefont {A.}~\bibnamefont {Vishwanath}},\ and\ \bibinfo {author} {\bibfnamefont {M.~P.~A.}\ \bibnamefont {Fisher}},\ }\bibfield  {title} {\bibinfo {title} {Quantum criticality beyond the landau-ginzburg-wilson paradigm},\ }\href {https://doi.org/10.1103/PhysRevB.70.144407} {\bibfield  {journal} {\bibinfo  {journal} {Phys. Rev. B}\ }\textbf {\bibinfo {volume} {70}},\ \bibinfo {pages} {144407} (\bibinfo {year} {2004})}\BibitemShut {NoStop}%
\bibitem [{\citenamefont {Fradkin}\ \emph {et~al.}(2004)\citenamefont {Fradkin}, \citenamefont {Huse}, \citenamefont {Moessner}, \citenamefont {Oganesyan},\ and\ \citenamefont {Sondhi}}]{FradkinEA04}%
  \BibitemOpen
  \bibfield  {author} {\bibinfo {author} {\bibfnamefont {E.}~\bibnamefont {Fradkin}}, \bibinfo {author} {\bibfnamefont {D.~A.}\ \bibnamefont {Huse}}, \bibinfo {author} {\bibfnamefont {R.}~\bibnamefont {Moessner}}, \bibinfo {author} {\bibfnamefont {V.}~\bibnamefont {Oganesyan}},\ and\ \bibinfo {author} {\bibfnamefont {S.~L.}\ \bibnamefont {Sondhi}},\ }\bibfield  {title} {\bibinfo {title} {Bipartite rokhsar--kivelson points and cantor deconfinement},\ }\href {https://doi.org/10.1103/PhysRevB.69.224415} {\bibfield  {journal} {\bibinfo  {journal} {Phys. Rev. B}\ }\textbf {\bibinfo {volume} {69}},\ \bibinfo {pages} {224415} (\bibinfo {year} {2004})}\BibitemShut {NoStop}%
\bibitem [{\citenamefont {Batista}\ and\ \citenamefont {Trugman}(2004)}]{BatistaTrugman04}%
  \BibitemOpen
  \bibfield  {author} {\bibinfo {author} {\bibfnamefont {C.~D.}\ \bibnamefont {Batista}}\ and\ \bibinfo {author} {\bibfnamefont {S.~A.}\ \bibnamefont {Trugman}},\ }\bibfield  {title} {\bibinfo {title} {{Exact Ground States of a Frustrated 2D Magnet: Deconfined Fractional Excitations at a First-Order Quantum Phase Transition}},\ }\href {https://doi.org/10.1103/PhysRevLett.93.217202} {\bibfield  {journal} {\bibinfo  {journal} {Phys. Rev. Lett.}\ }\textbf {\bibinfo {volume} {93}},\ \bibinfo {pages} {217202} (\bibinfo {year} {2004})}\BibitemShut {NoStop}%
\bibitem [{\citenamefont {Fradkin}(1991)}]{Fradkin}%
  \BibitemOpen
  \bibfield  {author} {\bibinfo {author} {\bibfnamefont {E.}~\bibnamefont {Fradkin}},\ }\href@noop {} {\emph {\bibinfo {title} {{Field theories of condensed matter systems}}}}\ (\bibinfo  {publisher} {Perseus Books (Reading, Massachusetts)},\ \bibinfo {year} {1991})\BibitemShut {NoStop}%
\bibitem [{\citenamefont {Polyakov}(1977)}]{Polyakov1977}%
  \BibitemOpen
  \bibfield  {author} {\bibinfo {author} {\bibfnamefont {A.}~\bibnamefont {Polyakov}},\ }\bibfield  {title} {\bibinfo {title} {{Quark confinement and topology of gauge theories}},\ }\href {https://doi.org/https://doi.org/10.1016/0550-3213(77)90086-4} {\bibfield  {journal} {\bibinfo  {journal} {Nuclear Physics B}\ }\textbf {\bibinfo {volume} {120}},\ \bibinfo {pages} {429 } (\bibinfo {year} {1977})}\BibitemShut {NoStop}%
\bibitem [{\citenamefont {Moessner}\ and\ \citenamefont {Sondhi}(2001)}]{MoessnerSondhi01}%
  \BibitemOpen
  \bibfield  {author} {\bibinfo {author} {\bibfnamefont {R.}~\bibnamefont {Moessner}}\ and\ \bibinfo {author} {\bibfnamefont {S.~L.}\ \bibnamefont {Sondhi}},\ }\bibfield  {title} {\bibinfo {title} {{Resonating Valence Bond Phase in the Triangular Lattice Quantum Dimer Model}},\ }\href@noop {} {\bibfield  {journal} {\bibinfo  {journal} {Phys. Rev. Lett.}\ }\textbf {\bibinfo {volume} {86}},\ \bibinfo {pages} {1881} (\bibinfo {year} {2001})}\BibitemShut {NoStop}%
\bibitem [{\citenamefont {Yan}\ \emph {et~al.}(2022)\citenamefont {Yan}, \citenamefont {Samajdar}, \citenamefont {Wang}, \citenamefont {Sachdev},\ and\ \citenamefont {Meng}}]{ZhengEA22}%
  \BibitemOpen
  \bibfield  {author} {\bibinfo {author} {\bibfnamefont {Z.}~\bibnamefont {Yan}}, \bibinfo {author} {\bibfnamefont {R.}~\bibnamefont {Samajdar}}, \bibinfo {author} {\bibfnamefont {Y.-C.}\ \bibnamefont {Wang}}, \bibinfo {author} {\bibfnamefont {S.}~\bibnamefont {Sachdev}},\ and\ \bibinfo {author} {\bibfnamefont {Z.~Y.}\ \bibnamefont {Meng}},\ }\bibfield  {title} {\bibinfo {title} {{Triangular lattice quantum dimer model with variable dimer density}},\ }\href {https://doi.org/10.1038/s41467-022-33431-5} {\bibfield  {journal} {\bibinfo  {journal} {Nature Communications}\ }\textbf {\bibinfo {volume} {13}},\ \bibinfo {pages} {5799} (\bibinfo {year} {2022})}\BibitemShut {NoStop}%
\bibitem [{\citenamefont {Sachdev}(1992)}]{Sachdev92}%
  \BibitemOpen
  \bibfield  {author} {\bibinfo {author} {\bibfnamefont {S.}~\bibnamefont {Sachdev}},\ }\bibfield  {title} {\bibinfo {title} {{Kagome and triangular-lattice Heisenberg antiferromagnets: Ordering from quantum fluctuations and quantum-disordered ground states with unconfined bosonic spinons}},\ }\href {https://doi.org/10.1103/PhysRevB.45.12377} {\bibfield  {journal} {\bibinfo  {journal} {Phys. Rev. B}\ }\textbf {\bibinfo {volume} {45}},\ \bibinfo {pages} {12377} (\bibinfo {year} {1992})}\BibitemShut {NoStop}%
\bibitem [{\citenamefont {Misguich}\ \emph {et~al.}(2002)\citenamefont {Misguich}, \citenamefont {Serban},\ and\ \citenamefont {Pasquier}}]{Misguich02}%
  \BibitemOpen
  \bibfield  {author} {\bibinfo {author} {\bibfnamefont {G.}~\bibnamefont {Misguich}}, \bibinfo {author} {\bibfnamefont {D.}~\bibnamefont {Serban}},\ and\ \bibinfo {author} {\bibfnamefont {V.}~\bibnamefont {Pasquier}},\ }\bibfield  {title} {\bibinfo {title} {{Quantum Dimer Model on the Kagome Lattice: Solvable Dimer-Liquid and Ising Gauge Theory}},\ }\href {https://doi.org/10.1103/PhysRevLett.89.137202} {\bibfield  {journal} {\bibinfo  {journal} {Phys. Rev. Lett.}\ }\textbf {\bibinfo {volume} {89}},\ \bibinfo {pages} {137202} (\bibinfo {year} {2002})}\BibitemShut {NoStop}%
\end{thebibliography}%

%
\appendix
\counterwithin{figure}{section}
\section{Appendix A: Free energy per dimer -- analytics}
\label{app:appendixA}

%

In this Appendix we calculate the free energy $f_{N}$ per dimer for the classical dimer model in an $N$-vertex region generated by $\mathcal{N}$ compositions of a spectre tile. The partition function is
\begin{equation}
\mathcal{Z}=2^{N_{\text{Mystic}}+1}
\end{equation}
and 
\begin{align}
f_{N} & =\frac{\ln\left(\mathcal{Z}\right)}{N/2}\\
 & =\frac{N_{\text{Mystic}}+1}{N/2}\cdot\ln\left(2\right).
\end{align}

In the bulk of any spectre tiling there are 6 vertices per spectre, found by summing the 14 vertices weighted by the internal angle the enclosed at each vertex. The number of spectres in any region is $N_{\text{Spectre}}+2N_{\text{Mystic}}$, since the Mystic contains two spectres. Therefore
\begin{equation}
N=6\left(N_{\text{Spectre}}+2N_{\text{Mystic}}\right)
\end{equation}
in the bulk. Using this expression for \emph{all} vertices will undercount
the vertices on the boundary, but this error will tend to zero in
the thermodynamic limit. The result is
\begin{equation}
f_{N}=\frac{N_{\text{Mystic}}+1}{N_{\text{Spectre}}+2N_{\text{Mystic}}}\cdot\frac{\ln\left(2\right)}{3}.\label{eq:fN}
\end{equation}

Fig.~\ref{fig:implied} shows the composition rules for the tiles.
Counting the tiles gives the following rule for the number of each
tile type after $\mathcal{N}$ compositions starting from a single
spectre:
\begin{equation}
\left(\begin{array}{c}
N_{\text{Spectre}}\left(\mathcal{N}\right)\\
N_{\text{Mystic}}\left(\mathcal{N}\right)
\end{array}\right)=\left(\begin{array}{cc}
7 & 6\\
1 & 1
\end{array}\right)^{\mathcal{N}}\left(\begin{array}{c}
1\\
0
\end{array}\right).\label{eq:matrix}
\end{equation}
The eigenvalues of this matrix are
\begin{equation}
\lambda_{\pm}=4\pm\sqrt{15}.
\end{equation}

To approach the thermodynamic limit $\mathcal{N}\rightarrow\infty$
systematically, note that the eigenvalues obey
\begin{equation}
\lambda^{2}=8\lambda-1
\end{equation}
from which it follows that 
\begin{align}
\lambda^{3} & =63\lambda-8\\
\lambda^{4} & =496\lambda-63\\
\lambda^{5} & =3905\lambda-496
\end{align}
and so on. In general
\begin{equation}
\lambda^{n}=L_{n}\lambda-L_{n-1}
\end{equation}
where $L_{0}=1$, $L_{1}=8$. This generates OEIS A001090:
\begin{equation}
L_{n}=1,\text{\,\ensuremath{\,}}8,\,\,63,\,\,496,\,\,3905,\,\,30744\ldots
\end{equation}
The first few terms in the series of Eq.~\ref{eq:matrix} are 
\begin{equation}
\left(\begin{array}{c}
N_{\text{Spectre}}\\
N_{\text{Mystic}}
\end{array}\right)=\left(\begin{array}{c}
1\\
0
\end{array}\right),\left(\begin{array}{c}
7\\
1
\end{array}\right),\left(\begin{array}{c}
55\\
8
\end{array}\right),\left(\begin{array}{c}
433\\
63
\end{array}\right),\ldots
\end{equation}
from which we see that 
\begin{align}
N_{\text{Spectre}}\left(\mathcal{N}\right)+N_{\text{Mystic}}\left(\mathcal{N}\right) & =L_{\mathcal{N}}\\
N_{\text{Mystic}}\left(\mathcal{N}\right) & =L_{\mathcal{N}-1}.
\end{align}
Substituting into Eq.~\ref{eq:fN} gives
\begin{align}
f_{N\left(\mathcal{N}\right)} & =\frac{L_{\mathcal{N}-1}+1}{L_{\mathcal{N}}+L_{\mathcal{N}-1}}\cdot\frac{\ln\left(2\right)}{3}.
\end{align}

Since $\lambda_{+}>1$ and $\left|\lambda_{-}\right|<1$ it follows
that $\lambda_{+}$ is a Pisot Vijayaraghavan number. The significance
is that in the thermodynamic limit $\mathcal{N}\rightarrow\infty$
only $\lambda_{+}$ will survive:
\begin{equation}
\lim_{\mathcal{N}\rightarrow\infty}L_{\mathcal{N}}/L_{\mathcal{N}-1}=\lambda_{+}
\end{equation}
and so 
\begin{equation}
\lim_{N\rightarrow\infty}f_{N}=\frac{1}{5+\sqrt{15}}\cdot\frac{\ln\left(2\right)}{3}.
\end{equation}

\section{Appendix B: Free energy per dimer -- numerics}
\label{app:appendixB}

Our raw data are given in Table \ref{tab:Raw-numerical-data}. Note
that the final row in the column $f_{N}$ is partly analytical: it
assumes Eq.~\eqref{eq:Z} for $\mathcal{Z}$ in the main text, but uses the numerical
value of $N$. Obtaining this value numerically requires finding the
determinant of an $1665406\times1665406$ matrix.

\begin{table}
\begin{raggedright}
\begin{tabular}{|c|c|c|c|c|}
\hline 
$N$ & $N/N_{T}$ & $f_{N}$ & Aitken $N/N_{T}$ & Aitken $f_{N}$\tabularnewline
\hline 
\hline 
78 & 8.666667 & 0.035546 &  & \tabularnewline
\hline 
518 & 7.295775 & 0.024086 & 6.177147 & 0.028448\tabularnewline
\hline 
3734 & 6.679785 & 0.023761 & 6.029943 & 0.026321\tabularnewline
\hline 
28006 & 6.363554 & 0.024601 & 6.004131 & 0.026074\tabularnewline
\hline 
214662 & 6.195330 & 0.025225 & 6.000536 & 0.026044\tabularnewline
\hline 
1665406 & 6.105062 & 0.025592{*} &  & \tabularnewline
\hline 
\end{tabular}
\par\end{raggedright}
\caption{\label{tab:Raw-numerical-data}Raw numerical data. $N$ is the number
of vertices, $N_{T}$ the number of $S_{0}$ spectre tiles, and $f_{N}$
the free energy per dimer. Aitken $N/N_{T}$ has had the Aitken delta
squared process applied to $N/N_{T}$ for faster convergence, and the
final column recalculates $f_{N}$ using the new $N/N_{T}$. Result
{*} is partly analytical (see text).}
\end{table}

It is evident from here and Fig.~\ref{fig:data} in the main text that the free
energy per dimer $f_{N}$ 
\begin{equation}
f_{N}=\frac{2\ln\left(\mathcal{Z}\right)}{N}
\end{equation}
is converging incredibly slowly. The issue is that the boundary of
the $S_{\mathcal{N}}$ region becomes highly fractal for large $\mathcal{N}$
(with a fractal dimension of around 1.44), so that a larger than expected
number of vertices lie on the boundary. One way to see the effect
is to find $N/N_{T}$, the number of vertices $N$ divided by the
number of $S_{0}$ spectre tiles $N_{T}$. As argued above, this value
should be 6 in the bulk, and should therefore tend to 6 as $\mathcal{N}\rightarrow\infty$.
Inspecting Table \ref{tab:Raw-numerical-data} shows a poor convergence.
To help with this we apply a nonlinear series acceleration, Aitken's
delta squared process~\cite{Aitken26}, in which a series $A_{n}$ is transformed
to a new series $A_{n}'$ with faster convergence:

\begin{equation}
A_{n}'=\frac{A_{n+1}A_{n-1}-A_{n}^{2}}{A_{n+1}+A_{n-1}-2A_{n}}.
\end{equation}

Table \ref{tab:Raw-numerical-data} shows the result of this transformation,
which is highly effective. We then re-calculate $f_{N}$ using 
\begin{equation}
f_{N}=\frac{2\ln\left(\mathcal{Z}\right)}{(N/N_{T})N_{T}}
\end{equation}
and the accelerated $N/N_{T}$ terms. The result is a quick convergence
to the analytical value.

\end{document}